# Quantitative analysis of secondary Bjerkenes forces in various liquids

Ion Simaciu [a,*], Zoltan Borsos [b,**], Viorel Drafta [c] and Gheorghe Dumitrescu [d]

[a] Retired lecturer, Petroleum-Gas University of Ploiești, Ploiești 100680, Romania

[b] Petroleum-Gas University of Ploiești, Ploiești 100680, Romania

[c] Independent researcher

[d] High School Toma N. Socolescu, Ploiești, Romania

E-mail: [*] isimaciu@yahoo.com ; [**] borzolh@upg-ploiesti.ro

**Abstract**
Numerically calculating the interaction forces between two free bubbles under the action of a background of random acoustic radiation, we highlight the contributions of coefficients $\beta_{s0}$ and $\beta_{a0} = \beta_{\mu 0} + \beta_{th0}$ to the size of these forces. The quantitative study of the forces is done for several fluids, as different as possible in terms of properties: water, mercury, liquid helium and superfluid helium. It is quantitatively demonstrated, for different radii of the oscillating bubbles, $R_0 = [10^{-1} - 10^{-10}]$m, that the scattering–absorption forces and the scattering–scattering forces are close in magnitude. For: water, mercury and liquid helium, the force ratio is in the range $f_a \in (10^{-3} - 1)$. For superfluid helium, the forces change direction, oscillatingly, and the ratio of the forces is much less than one, $|f_a| << 1$.

## 1. Introduction

In the previously published papers [1-4], I tried to identify, by analogy with the electromagnetic world, the electrostatic acoustic forces and the gravitational acoustic forces for two free bubbles. Based on the fact that, in the electromagnetic world [5- pp. 801-806], the absorption/attenuation coefficient for the electromagnetic oscillator (the absorptive width $\Gamma_i = \tau_i \omega_{0i}^2 = 2R_i \omega_{0i}^2/(3c) = 2n_i^2 e^2 \omega_{0i}^2/(3m_i c^3)$) is much smaller than the radiative coefficient (the radiative decay - in acoustic terminology, acoustic damping coefficient) and therefore implicitly the radiation scattering section is much larger than the absorption section, we considered that the gravitational acoustic forces are the ones that involve scattering and absorption [3]. This hypothesis is also supported by the results of the modelling of the electrostatic interaction and the gravitational interaction in Stochastic Electrodynamics [6, 7]. In what it follow we will show that, for real liquids: water, mercury and liquid and superfluid helium, the absorption damping coefficient (viscous $\beta_{\mu 0}$ and thermal $\beta_{th0}$, $\beta_{a0} = \beta_{\mu 0} + \beta_{th0}$), for certain values of the radii, $R_0 = [10^{-1} - 10^{-10}]$m, has an as large contribution to the Bjerkenes force, in a radiation background, as the acoustic coefficient $\beta_{s0}$ (the zero index denotes the value of the coefficient for natural pulsation $\omega_0$, $\beta_0 = \beta(\omega_0)$).

For superfluid helium ($T \in [1.5 - 2.16]$K), the results of the calculation of the acoustic force, as a function of the bubble radius, highlight a strong dependence on the diffusion coefficient that leads to variations between positive and negative values (change of sign and therefore meaning) of the acoustic forces for close values of bubbles radii. This result is explained by the dependence of the thermal absorption coefficient, $\beta_{th0}$, on the trigonometric functions sin and cos. Also, the values of the coefficient $\beta_{th0}$ are greater, in absolute value, than the coefficients, $\beta_{s0}$ and $\beta_{\mu 0}$ so their sum can be negative. We chose these liquids for their properties: water being the most common liquid on earth, mercury as the only liquid metal at room temperature, and helium because it can be both liquid and superfluid.

In a future paper we will demonstrate that the scattering-absorption forces between two bubbles in a cluster [8], in a radiation background, are much smaller than the scattering-scattering forces, if the number of bubbles is large.

In the second section we calculate the acoustic forces and implicitly the square of the acoustic charge for real liquids: 2.1. Bubbles in water, 2.2. Bubbles in mercury, 2.3. Bubbles in liquid helium, 2.4. Bubbles in superfluid helium. In the third section we calculate the ratio between the scattering-absorption forces and the scattering-scattering forces, $f_a = F_{BsaT}/F_{BssT} = F_{BsaZ}/F_{BssZ}$, in the radiation background, for three fluids: water, mercury and helium. Calculations show that these forces are close in magnitude for the same bubble radii. The exception is this ratio calculated for bubbles in superfluid helium, which is much smaller than one. In the fourth section we analyze the results obtained and present the conclusions.

## 2. Calculation of acoustic forces for real liquids

### 2.1. Bubbles in water

According to Eqs. (71) from the paper [1], the expressions of the Bjerkenes force between two free bubbles in acoustic radiation background are:

$$F_{BT}(r) = \left(\frac{-\hbar u}{r^2}\right) \frac{R_0 \omega_0}{u\left(\exp\frac{\hbar \omega_0}{kT} - 1\right)\left(1 + \frac{4\mu u}{\rho R_0^3 \omega_0^2} + \frac{2u\beta_{th0}}{R_0 \omega_0^2}\right)},$$

$$F_{BZ}(r) = \left(\frac{-\hbar u}{r^2}\right) \frac{R_0 \omega_0}{2u\left(1 + \frac{4\mu u}{\rho R_0^3 \omega_0^2} + \frac{2u\beta_{th0}}{R_0 \omega_0^2}\right)}.$$
(1)

with: $\omega_0^2 = p_{eff}/(\rho R_0^2) = [3\gamma p_0 + 2(3\gamma - 1)\sigma/R_0]/(\rho R_0^2) = [3\gamma p_0 R_0 + 2(3\gamma - 1)\sigma]/(\rho R_0^3)$,

$$\beta_{th0} = \frac{3(\gamma - 1)\left[X_0(\sinh X_0 + \sin X_0) - 2(\cosh X_0 - \cos X_0)\right]\omega_0}{2X_0\left[X_0(\cosh X_0 - \cos X_0) + 3(\gamma - 1)(\sinh X_0 - \sin X_0)\right]}$$
(2)

and $X_0 = R_0(2\omega_0/\chi)^{1/2}$.

By analogy with the electromagnetic world, in the papers published so far [1-4], we wrongly considered that also in the acoustic world the absorption damping coefficient, $\beta_{a0} = \beta_{\mu 0} + \beta_{th0}$, is much lower than the acoustic re-radiated/ scattered component, $\beta_{s0} = \omega_0^2 R_0/(2u)$. Next we will show that we cannot make this approximation because, for certain values of the radii, these coefficients have close contributions. We will do the calculations for the interaction between bubbles in some real liquids.

To highlight the contribution of the scattering and absorption phenomena in the magnitudes of the forces, we do the numerical evaluation of these forces, for the liquid water, for the range of sizes of the bubble radii $R_0 = [10^{-1} - 10^{-10}]$m. Water parameters, relevant for the phenomenon of interaction between bubbles and air, are: $p_0 = 10^5$Pa static pressure in liquid, $\mu = 10^{-3}$Pas dynamic viscosity, $\sigma = 0.0725$Nm$^{-1}$ surface tension, $u = 1.510^3$ms$^{-1}$ sound velocity, $\gamma = 1.4$ polytropic exponents of the air in the bubble, $\rho = 10^3$kgm$^{-3}$ mass density, $\chi = 1.14 \cdot 10^{-4}$m$^2$s$^{-1}$ self-diffusion coefficient for air, $T = 300$K liquid temperature and the universal constants $k = 1.381 \times 10^{-23}$ J·K$^{-1}$, $\hbar = 1.05 \times 10^{-34}$Js [9]. The results are presented in tables T 1 and T 2. Table T 1 contains the ratio values: $\hbar\omega_0/(kT)$, $4\mu u/(\rho R_0^3 \omega_0^2)$, $2u\beta_{th0}/(R_0\omega_0^2)$, $\alpha_{aZ}$ and $\alpha_{aZ}$.

| $R_0[m]$ | $\dfrac{\hbar\omega_0}{kT}$ | $\dfrac{4u\mu}{\rho R^3 \omega_0^2}$ | $\dfrac{2u\beta_{tho}}{R\omega_0^2}$ | $\alpha_{aZ} = \dfrac{e_{aZ}^2(R_0)}{\hbar u}$ | $\alpha_{aT} = \dfrac{e_{aT}^2(R_0)}{\hbar u}$ |
|---|---|---|---|---|---|
| $10^{-1}$ | $5.198 \cdot 10^{-12}$ | $1.429 \cdot 10^{-4}$ | $4.554 \cdot 10^{-1}$ | $4.693 \cdot 10^{-3}$ | $2.551 \cdot 10^{-2}$ |
| $10^{-2}$ | $5.198 \cdot 10^{-11}$ | $1.428 \cdot 10^{-3}$ | $1.388$ | $2.859 \cdot 10^{-3}$ | $1.554 \cdot 10^{-2}$ |
| $10^{-3}$ | $5.201 \cdot 10^{-10}$ | $1.427 \cdot 10^{-2}$ | $3.894$ | $1.393 \cdot 10^{-3}$ | $7.571 \cdot 10^{-3}$ |
| $10^{-4}$ | $5.226 \cdot 10^{-9}$ | $1.413 \cdot 10^{-1}$ | $8.115$ | $7.42 \cdot 10^{-4}$ | $4.034 \cdot 10^{-3}$ |
| $10^{-5}$ | $5.477 \cdot 10^{-8}$ | $1.286$ | $2.439$ | $1.523 \cdot 10^{-3}$ | $8.282 \cdot 10^{-3}$ |
| $10^{-6}$ | $7.541 \cdot 10^{-7}$ | $6.787$ | $2.505 \cdot 10^{-1}$ | $1.233 \cdot 10^{-3}$ | $6.703 \cdot 10^{-3}$ |
| $10^{-7}$ | $1.804 \cdot 10^{-5}$ | $1.186 \cdot 10^{1}$ | $2.506 \cdot 10^{-2}$ | $1.841 \cdot 10^{-3}$ | $1.001 \cdot 10^{-2}$ |
| $10^{-8}$ | $5.488 \cdot 10^{-4}$ | $1.282 \cdot 10^{1}$ | $2.506 \cdot 10^{-3}$ | $5.22 \cdot 10^{-3}$ | $2.836 \cdot 10^{-2}$ |
| $10^{-9}$ | $1.728 \cdot 10^{-2}$ | $1.292 \cdot 10^{1}$ | $2.506 \cdot 10^{-4}$ | $1.632 \cdot 10^{-2}$ | $8.72 \cdot 10^{-2}$ |
| $10^{-10}$ | $5.463 \cdot 10^{-1}$ | $1.293 \cdot 10^{1}$ | $2.506 \cdot 10^{-5}$ | $5.155 \cdot 10^{-2}$ | $1.623 \cdot 10^{-1}$ |

T 1

The constants $\alpha_{aZ}$ and $\alpha_{aZ}$ have the meaning of fine acoustic constants, i.e. they are the ratio between the square of the acoustic charges $e_a^2$ and the square of the natural acoustic charge $\hbar u$ $\left(\alpha_a = e_a^2/(\hbar u)\right)$. In the electromagnetic world, the fine structure constant is the ratio between the square of the electrostatic charge of the electron and the square of the natural electric charge (it is considered to be the maximum charge in the acoustic world) $\alpha = e^2/(\hbar c) \cong 1/137 \cong 7.3 \cdot 10^{-3}$ [5 - Subch. 6.12]. As can be seen from table 2, the square of the acoustic charge is much smaller than the square of the electrostatic charge $e_a^2 << e^2 = q_e^2/(4\pi\varepsilon_0) \cong 2.2 \cdot 10^{-28} \text{Nm}^2 = 2.2 \cdot 10^{-28} \text{kgm}^3\text{s}^{-2}$, for any radius in the range $R_0 = \left[10^{-1} - 10^{-10}\right]$m.

According to this table, the ratio $x = \hbar\omega_0/(kT)$ is much less than one for all values of the bubble equilibrium radius, and so we can approximate $\exp x - 1 \cong x$. With this approximation, the expression of the force in the thermal background becomes

$$F_{BT}(r) = \left(\frac{-\hbar u}{r^2}\right) \frac{R_0 \omega_0}{u\left(\exp\dfrac{\hbar\omega_0}{kT} - 1\right)\left(1 + \dfrac{4\mu u}{\rho R_0^3 \omega_0^2} + \dfrac{2u\beta_{th0}}{R_0\omega_0^2}\right)} \cong \left(\frac{-kTR_0}{r^2}\right) \frac{1}{\left(1 + \dfrac{4\mu u}{\rho R_0^3 \omega_0^2} + \dfrac{2u\beta_{th0}}{R_0\omega_0^2}\right)}. \quad (3)$$

Also, since for small radii, $R_0 = \left[10^{-6} - 10^{-10}\right]$m, the contribution of thermal effects is much smaller than the viscous ones, $2u\beta_{th0}/(R_0\omega_0^2) << 4\mu u/(\rho R_0^3 \omega_0^2)$, we can analytically approximate the force expression according to the relation

$$F_{BT}(r) \cong \left(\frac{-kTR_0}{r^2}\right)\frac{1}{\left(1+\dfrac{4\mu u}{\rho R_0^3 \omega_0^2}\right)} = \left(\frac{-kTR_0}{r^2}\right)\frac{\rho R_0^3 \omega_0^2}{4\mu u\left(1+\dfrac{\rho R_0^3 \omega_0^2}{4\mu u}\right)} \cong \left(\frac{-kTR_0}{r^2}\right)\frac{\rho R_0^3 \omega_0^2}{4\mu u}$$

$$= \left(\frac{-\sigma_0^2}{r^2}\right)\frac{\rho \omega_0^2 kT}{64\pi^2 \mu u} = \left(\frac{-\sigma_0^2}{r^2}\right)\frac{kT}{64\pi^2 R_0^2 \mu u} = \left(\frac{-\sigma_0^2}{r^2}\right)w_{T0}. \quad \sigma_0^2 = 4\pi R_0^2, w_{T0} = \frac{kT}{64\pi^2 R_0^2 \mu u}. \quad (4)$$

The same relation (3), for large radii $R_0 = \left[10^{-4} - 10^{-1}\right]$m, since $2u\beta_{th0}/(R_0\omega_0^2) >> 4\mu u/(\rho R_0^3 \omega_0^2)$, can be approximated analytically by

$$F_{BT}(r) = \left(\frac{-kTR_0}{r^2}\right)\frac{1}{\left(1+\dfrac{2u\beta_{th0}}{R_0\omega_0^2}\right)} \cong \left(\frac{-kTR_0}{r^2}\right)\frac{R_0 \omega_0^2}{2u\beta_{th0}\left(1+\dfrac{R_0\omega_0^2}{2u\beta_{th0}}\right)} \cong \left(\frac{-kTR_0}{r^2}\right)\left(\frac{R_0\omega_0^2}{2u\beta_{th0}}\right). \quad (5)$$

Also, for these radii ($R_0 = [10^{-4} - 10^{-1}]$m), according to table T 2,

| $R_0$[m] | $\omega_0$[s$^{-1}$] | $X_0$ | $\beta_{th0}$[s$^{-1}$] | $\dfrac{R_0\omega_0}{u}$ | $e_{aZ}^2$[Nm$^2$] | $e_{aT}^2$[Nm$^2$] |
|---|---|---|---|---|---|---|
| $10^{-1}$ | $2.049 \cdot 10^2$ | $1.896 \cdot 10^2$ | $6.376 \cdot 10^{-1}$ | $1.366 \cdot 10^{-2}$ | $7.392 \cdot 10^{-34}$ | $4.019 \cdot 10^{-33}$ |
| $10^{-2}$ | $2.05 \cdot 10^3$ | $5.996 \cdot 10^1$ | $1.943 \cdot 10^1$ | $1.366 \cdot 10^{-2}$ | $4.503 \cdot 10^{-34}$ | $2.448 \cdot 10^{-33}$ |
| $10^{-3}$ | $2.051 \cdot 10^4$ | $1.897 \cdot 10^1$ | $5.457 \cdot 10^2$ | $1.367 \cdot 10^{-2}$ | $2.193 \cdot 10^{-34}$ | $1.192 \cdot 10^{-33}$ |
| $10^{-4}$ | $2.061 \cdot 10^5$ | $6.013$ | $1.149 \cdot 10^4$ | $1.374 \cdot 10^{-2}$ | $1.169 \cdot 10^{-34}$ | $6.354 \cdot 10^{-34}$ |
| $10^{-5}$ | $2.16 \cdot 10^6$ | $1.946$ | $3.792 \cdot 10^4$ | $1.44 \cdot 10^{-2}$ | $2.399 \cdot 10^{-34}$ | $1.304 \cdot 10^{-33}$ |
| $10^{-6}$ | $2.973 \cdot 10^7$ | $7.222 \cdot 10^{-1}$ | $7.381 \cdot 10^4$ | $1.982 \cdot 10^{-2}$ | $1.942 \cdot 10^{-34}$ | $1.056 \cdot 10^{-33}$ |
| $10^{-7}$ | $7.113 \cdot 10^8$ | $3.533 \cdot 10^{-1}$ | $4.227 \cdot 10^5$ | $4.742 \cdot 10^{-2}$ | $2.899 \cdot 10^{-34}$ | $1.576 \cdot 10^{-33}$ |
| $10^{-8}$ | $2.164 \cdot 10^{10}$ | $1.948 \cdot 10^{-1}$ | $3.911 \cdot 10^6$ | $1.443 \cdot 10^{-1}$ | $8.221 \cdot 10^{-34}$ | $4.467 \cdot 10^{-33}$ |
| $10^{-9}$ | $6.815 \cdot 10^{11}$ | $1.093 \cdot 10^{-1}$ | $3.88 \cdot 10^7$ | $4.543 \cdot 10^{-1}$ | $2.57 \cdot 10^{-33}$ | $1.373 \cdot 10^{-32}$ |
| $10^{-10}$ | $2.154 \cdot 10^{13}$ | $6.148 \cdot 10^{-2}$ | $3.877 \cdot 10^8$ | $1.436$ | $8.119 \cdot 10^{-33}$ | $2.556 \cdot 10^{-32}$ |

T2

parameter $X_0$ is $X_0 = R_0(2\omega_0/\chi)^{1/2} \gg 1$, the coefficient $\beta_{th0}$ can be approximated, according to Eq. (17a) from paper [2] with $\beta_{th0} = 3(\gamma-1)\sqrt{\chi\omega_0}/(2\sqrt{2}R_0)$. Inserting this expression into Eq. (15), it follows

$$F_{BT}(r) \cong \left(\frac{-kTR_0}{r^2}\right)\left(\frac{R_0\omega_0^2\sqrt{2}R_0}{3(\gamma-1)\sqrt{\chi\omega_0}u}\right) = \left(\frac{-\sigma_0^2}{r^2}\right)\left(\frac{kT\omega_0}{24(\gamma-1)\pi R_0^2 u}\sqrt{\frac{\omega_0}{2\chi}}\right) =$$
$$\left(\frac{-\sigma_0^2}{r^2}\right)w_{T\chi 0}, \quad w_{T\chi 0} = \frac{kT\omega_0}{24(\gamma-1)\pi^2 R_0 u}\sqrt{\frac{\omega_0}{2\chi}}.$$
(6)

Expressions (4, 6) of the force resemble the expression of the electrostatic force between two electrons, Eq. (30), from papers [1 – arxiv V2], written according to the Thomson cross section, $\sigma_T = (8\pi/3)(e^2/m_e c^2)^2$.

Using the same reasoning of analytical approximation, the expression of the force in the acoustic background CZPF, becomes: a) for small radii, $R_0 = [10^{-6} - 10^{-10}]$m,

$$F_{BZ}(r) = \left(\frac{-\hbar u}{r^2}\right)\frac{R_0\omega_0}{2u\left(1 + \dfrac{4\mu u}{\rho R_0^3 \omega_0^2} + \dfrac{2u\beta_{th0}}{R_0\omega_0^2}\right)} \cong \left(\frac{-\hbar u}{r^2}\right)\frac{R_0\omega_0}{2u\left(1 + \dfrac{4\mu u}{\rho R_0^3\omega_0^2}\right)} \cong \left(\frac{-\hbar u}{r^2}\right)\frac{\rho R_0^4\omega_0^3}{8\mu u^2\left(1 + \dfrac{\rho R_0^3\omega_0^2}{4\mu u}\right)} \cong$$
$$\left(\frac{-\hbar u}{r^2}\right)\frac{\rho R_0^4\omega_0^3}{8\mu u^2} = \left(\frac{-\sigma_0^2}{r^2}\right)\frac{\rho\hbar\omega_0^3}{128\pi^2\mu u} = \left(\frac{-\sigma_0^2}{r^2}\right)w_{0\mu}, \quad \sigma_0 = 4\pi R_0^2, \quad w_{0\mu} = \frac{\rho\hbar\omega_0^3}{128\pi^2\mu u}$$
(7)

and b) for large radii $R_0 = [10^{-4} - 10^{-2}]$m,

$$F_{BZ}(r) = \left(\frac{-\hbar u}{r^2}\right)\frac{R_0\omega_0}{2u\left(1 + \dfrac{4\mu u}{\rho R_0^3\omega_0^2} + \dfrac{2u\beta_{th0}}{R_0\omega_0^2}\right)} \cong \left(\frac{-\hbar u}{r^2}\right)\frac{R_0\omega_0}{2u\left(1 + \dfrac{2u\beta_{th0}}{R_0\omega_0^2}\right)} \cong$$
$$\left(\frac{-\hbar u}{r^2}\right)\frac{R_0^2\omega_0^3}{4u^2\beta_{th0}\left(1 + \dfrac{R_0\omega_0^2}{2u\beta_{th0}}\right)} \cong \left(\frac{-\hbar u}{r^2}\right)\frac{R_0^2\omega_0^3}{4u^2\beta_{th0}} = \left(\frac{-1}{r^2}\right)\frac{\hbar R_0^3\omega_0^3}{3\sqrt{2}u(\gamma-1)\sqrt{\chi\omega_0}} =$$
$$\left(\frac{-\sigma_0^2}{r^2}\right)\left(\frac{\hbar\omega_0^2}{48\sqrt{2}(\gamma-1)\pi^2 R_0 u}\sqrt{\frac{\omega_0}{\chi}}\right) = \left(\frac{-\sigma_0^2}{r^2}\right)w_{0\chi}, \quad \sigma_0 = 4\pi R_0^2,$$
(8)
$$w_{0\chi} = \frac{\hbar\omega_0^2}{48\sqrt{2}(\gamma-1)\pi^2 R_0 u}\sqrt{\frac{\omega_0}{\chi}}.$$

According to Eqs. (7, 8), also the expressions of this type of acoustic forces can be put in a similar form to the electrostatic force. We also highlight, according to the papers [1], an acoustic charge having the square $e_a^2$:

$$e_{aT}^2 = \frac{\hbar R_0 \omega_0}{\left(\exp\frac{\hbar\omega_0}{kT}-1\right)\left(1+\frac{4\mu u}{\rho R_0^3 \omega_0^2}+\frac{2u\beta_{th0}}{R_0 \omega_0^2}\right)} = (\hbar u)\alpha_{aT},$$

$$\alpha_{aT} = \frac{e_{aT}^2}{\hbar u} = \frac{R_0 \omega_0}{u\left(\exp\frac{\hbar\omega_0}{kT}-1\right)\left(1+\frac{4\mu u}{\rho R_0^3 \omega_0^2}+\frac{2u\beta_{th0}}{R_0 \omega_0^2}\right)}; \quad (9)$$

$$e_{aZ}^2 = \frac{\hbar R_0 \omega_0}{2\left(1+\frac{4\mu u}{\rho R_0^3 \omega_0^2}+\frac{2u\beta_{th0}}{R_0 \omega_0^2}\right)} = (\hbar u)\alpha_{aZ}, \quad \alpha_{aZ} = \frac{e_{aZ}^2}{\hbar u} = \frac{R_0 \omega_0}{2u\left(1+\frac{4\mu u}{\rho R_0^3 \omega_0^2}+\frac{2u\beta_{th0}}{R_0 \omega_0^2}\right)}.$$

According to table T 2 and graph (graphic representation) G 1, for radii around the value $R_0 = 5.5\cdot 10^{-5}$ m, the contributions to the value of the charge and therefore implicitly of the acoustic force of: the acoustic coefficient, $\beta_{s0} = \beta_s(\omega_0) = \omega_0^2 R_0/(2u)$, the viscous coefficient $\beta_\mu = \beta(\omega_0) = 2\mu/(\rho R_0^2)$ and the thermal coefficient $\beta_{th0}$ are close in value. Acoustic charges, fine acoustic constants and implicitly acoustic forces have a minimum value for two radii: $R_{w1} \cong 6,695\times 10^{-7}$ m and $R_{w2} \cong 7,370\times 10^{-5}$ m. The square of the acoustic charges corresponding to the two radii are: $e_{aZ1}^2 \cong 1,89\cdot 10^{-34}$ Nm², $e_{aT1}^2 \cong 1,03\cdot 10^{-33}$ Nm², $e_{aZ2}^2 \cong 1,14\cdot 10^{-34}$ Nm² and $e_{aT2}^2 \cong 6,21\cdot 10^{-34}$ Nm².

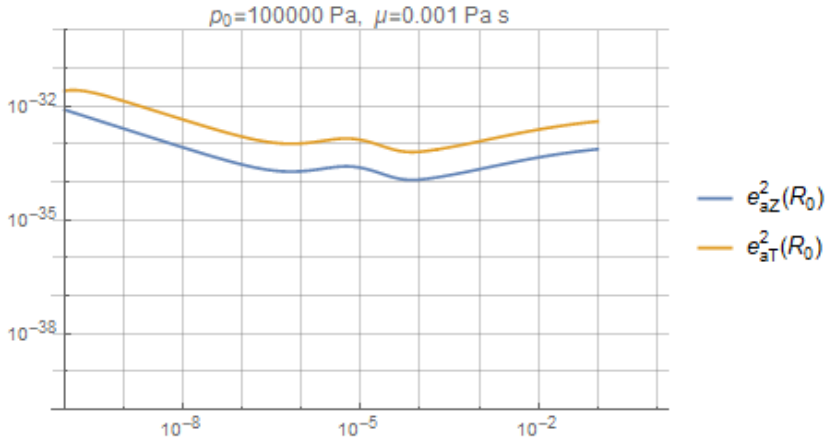

G 1

In the case when the experiments of measuring acoustic forces are performed in weightlessness (in a laboratory placed on a space station), the pressure in the enclosure is $p_0 = 0$ Pa. The study of acoustic forces in weightlessness conditions is motivated by the fact that, by canceling the effects of gravity, it is easier to observe, experimentally, the acoustic forces and therefore we can study the properties of these forces and the phenomena induced by them. In this case, the effective pressure is determined by the surface tension and has the expression $p_{eff} = 2(3\gamma-1)\sigma/R_0$ and the natural pulsation/ resonance pulsation has the expression $\omega_0^2 = 2(3\gamma-1)\sigma/(\rho R_0^3)$. Numerical results are given in tables T 3 and T 4 and graph G 2.

| $R_0$[m] | $\frac{\hbar\omega_0}{kT}$ | $\frac{4\mu u}{\rho R^3 \omega_0^2}$ | $\frac{2u\beta_{th0}}{R\omega_0^2}$ | $\alpha_{aZ} = \frac{e_{aZ}^2(R_0)}{\hbar u}$ | $\alpha_{aT} = \frac{e_{aT}^2(R_0)}{\hbar u}$ |
|---|---|---|---|---|---|
| $10^{-1}$ | $1.728\cdot 10^{-14}$ | $1.293\cdot 10^1$ | $1.78\cdot 10^3$ | $1.266\cdot 10^{-8}$ | $6.883\cdot 10^{-8}$ |
| $10^{-2}$ | $5.463\cdot 10^{-13}$ | $1.293\cdot 10^1$ | $7.709\cdot 10^2$ | $9.149\cdot 10^{-8}$ | $4.974\cdot 10^{-7}$ |
| $10^{-3}$ | $1.728\cdot 10^{-11}$ | $1.293\cdot 10^1$ | $1.976\cdot 10^2$ | $1.073\cdot 10^{-6}$ | $5.836\cdot 10^{-6}$ |
| $10^{-4}$ | $5.463\cdot 10^{-10}$ | $1.293\cdot 10^1$ | $2.439\cdot 10^1$ | $1.874\cdot 10^{-5}$ | $1.019\cdot 10^{-4}$ |
| $10^{-5}$ | $1.728\cdot 10^{-8}$ | $1.293\cdot 10^1$ | $2.499$ | $1.382\cdot 10^{-4}$ | $7.513\cdot 10^{-4}$ |
| $10^{-6}$ | $5.463\cdot 10^{-7}$ | $1.293\cdot 10^1$ | $2.506\cdot 10^{-1}$ | $5.063\cdot 10^{-4}$ | $2.753\cdot 10^{-3}$ |
| $10^{-7}$ | $1.728\cdot 10^{-5}$ | $1.293\cdot 10^1$ | $2.506\cdot 10^{-2}$ | $1.627\cdot 10^{-3}$ | $8.845\cdot 10^{-3}$ |
| $10^{-8}$ | $5.463\cdot 10^{-4}$ | $1.293\cdot 10^1$ | $2.506\cdot 10^{-3}$ | $5.153\cdot 10^{-3}$ | $2.8\cdot 10^{-2}$ |
| $10^{-9}$ | $1.728\cdot 10^{-2}$ | $1.293\cdot 10^1$ | $2.506\cdot 10^{-4}$ | $1.63\cdot 10^{-2}$ | $8.709\cdot 10^{-2}$ |
| $10^{-10}$ | $5.463\cdot 10^{-1}$ | $1.293\cdot 10^1$ | $2.506\cdot 10^{-5}$ | $5.154\cdot 10^{-2}$ | $1.623\cdot 10^{-1}$ |

T 3

According to table T 3, for radii in the interval, $R_0 = [10^{-1} - 10^{-4}]$ m, $2u\beta_{th0}/(R_0\omega_0^2) > 4\mu u/(\rho R_0^3 \omega_0^2) \gg 1$. For radii in the interval $R_0 = [10^{-5} - 10^{-10}]$ m, $2u\beta_{th0}/(R_0\omega_0^2) < 4\mu u/(\rho R_0^3 \omega_0^2) \ll 1$.

| $R_0$[m] | $\omega_0$[s$^{-1}$] | $X_0$ | $\beta_{th0}$[s$^{-1}$] | $\frac{R_0\omega_0}{u}$ | $e_{aZ}^2$[Nm$^2$] | $e_{aT}^2$[Nm$^2$] |
|---|---|---|---|---|---|---|
| $10^{-1}$ | $6.812 \cdot 10^{-1}$ | $1.093 \cdot 10^1$ | $2.752 \cdot 10^{-2}$ | $4.541 \cdot 10^{-5}$ | $1.994 \cdot 10^{-39}$ | $1.084 \cdot 10^{-38}$ |
| $10^{-2}$ | $2.154 \cdot 10^1$ | $6/147$ | $1.192$ | $1.436 \cdot 10^{-4}$ | $1.441 \cdot 10^{-38}$ | $7.834 \cdot 10^{-38}$ |
| $10^{-3}$ | $6.812 \cdot 10^2$ | $3.457$ | $3.056 \cdot 10^1$ | $4.541 \cdot 10^{-4}$ | $1.691 \cdot 10^{-37}$ | $9.191 \cdot 10^{-37}$ |
| $10^{-4}$ | $2.154 \cdot 10^4$ | $1.944$ | $3.773 \cdot 10^2$ | $1.436 \cdot 10^{-3}$ | $2.951 \cdot 10^{-36}$ | $1.604 \cdot 10^{-35}$ |
| $10^{-5}$ | $6.812 \cdot 10^5$ | $1.093$ | $3.866 \cdot 10^3$ | $4.541 \cdot 10^{-3}$ | $2.177 \cdot 10^{-35}$ | $1.183 \cdot 10^{-34}$ |
| $10^{-6}$ | $2.154 \cdot 10^7$ | $6.147 \cdot 10^{-1}$ | $3.875 \cdot 10^4$ | $1.436 \cdot 10^{-2}$ | $7.974 \cdot 10^{-35}$ | $4.335 \cdot 10^{-34}$ |
| $10^{-7}$ | $6.812 \cdot 10^8$ | $3.457 \cdot 10^{-1}$ | $3.876 \cdot 10^5$ | $4.541 \cdot 10^{-2}$ | $2.562 \cdot 10^{-34}$ | $1.393 \cdot 10^{-33}$ |
| $10^{-8}$ | $2.154 \cdot 10^{10}$ | $1.944 \cdot 10^{-1}$ | $3.876 \cdot 10^6$ | $1.436 \cdot 10^{-1}$ | $8.116 \cdot 10^{-34}$ | $4.41 \cdot 10^{-33}$ |
| $10^{-9}$ | $6.812 \cdot 10^{11}$ | $1.093 \cdot 10^{-1}$ | $3.876 \cdot 10^7$ | $4.541 \cdot 10^{-1}$ | $2.567 \cdot 10^{-33}$ | $1.372 \cdot 10^{-32}$ |
| $10^{-10}$ | $2.154 \cdot 10^{13}$ | $6.147 \cdot 10^{-2}$ | $3.876 \cdot 10^8$ | $1.436$ | $8.118 \cdot 10^{-33}$ | $2.556 \cdot 10^{-32}$ |

T 4

Comparing the results from tables T 2 and T 4, it follows that the square of the sound charge in water, for $p_0 = 10^5$ Pa, for large radii $R_0 = [10^{-1} - 10^{-5}]$ m, is greater than the square of the sound charge in water, for $p_0 = 0$ Pa and for the same radius. For small radii, $R_0 = [10^{-5} - 10^{-9}]$ m, the square of the acoustic charges, for the two pressures, $p_0$, become closer, for the same radius, so that for $R_0 = 10^{-10}$ m become almost equal.

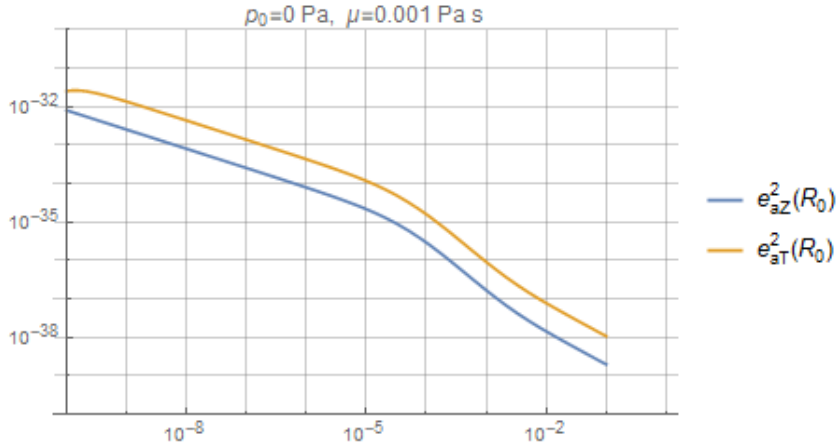

G 2

For bubbles in water in weightlessness, according to graph G 2, the acoustic fine structure constant, the square of the charge and implicitly the acoustic forces decrease with the increase of the bubble radius.

## 2.2. Bubbles in mercury

The parameters of mercury, relevant for the phenomenon of interaction between bubbles and air, are: $p_0 = 10^5$ Pa static pressure in liquid, $\mu = 1.4 \cdot 10^{-3}$ Pa s dynamic viscosity, $\sigma = 4.84 \cdot 10^{-1}$ Nm$^{-1}$ surface tension, $u = 1.45 10^3$ ms$^{-1}$ sound velocity, $\gamma = 1.4$ polytropic exponents of the air in the bubble, $\rho = 13.534 \cdot 10^3$ kgm$^{-3}$ mass density, $\chi = 4 \cdot 10^{-6}$ m$^2$s$^{-1}$ self-diffusion coefficient for air, $T = 300$ K liquid temperature and the universal constants $k = 1.381 \times 10^{-23}$ J·K$^{-1}$, $\hbar = 1.05 \times 10^{-34}$ Js [10]. The results are shown in tables T 5, T 6 and graph G 3.

|  | $R_0$ | $\frac{\hbar\omega_0}{kT}$ | $\frac{4u\mu}{\rho R^3 \omega_0^2}$ | $\frac{2u\beta_{th0}}{R\omega_0^2}$ | $\alpha_{aZ} = \frac{e_{aZ}^2(R_0)}{\hbar u}$ | $\alpha_{aT} = \frac{e_{aT}^2(R_0)}{\hbar u}$ |
|---|---|---|---|---|---|---|
| $p_0 = 10000$ Pa | $10^{-1}$ | $1.413 \cdot 10^{-12}$ | $1.933 \cdot 10^{-4}$ | $5.882 \cdot 10^{-1}$ | $1.209 \cdot 10^{-3}$ | $6.575 \cdot 10^{-3}$ |
| $\mu = 0.0014$ Pa s | $10^{-2}$ | $1.413 \cdot 10^{-11}$ | $1.932 \cdot 10^{-3}$ | $1.835$ | $6.774 \cdot 10^{-4}$ | $3.683 \cdot 10^{-3}$ |
|  | $10^{-3}$ | $1.418 \cdot 10^{-10}$ | $1.919 \cdot 10^{-2}$ | $5.537$ | $2.941 \cdot 10^{-4}$ | $1.599 \cdot 10^{-3}$ |
|  | $10^{-4}$ | $1.464 \cdot 10^{-9}$ | $1.801 \cdot 10^{-1}$ | $1.462 \cdot 10^{1}$ | $1.26 \cdot 10^{-4}$ | $6.849 \cdot 10^{-4}$ |
|  | $10^{-5}$ | $1.862 \cdot 10^{-8}$ | $1.113$ | $2.196 \cdot 10^{1}$ | $1.052 \cdot 10^{-4}$ | $5.718 \cdot 10^{-4}$ |
|  | $10^{-6}$ | $4.089 \cdot 10^{-7}$ | $2.308$ | $6.144$ | $5.881 \cdot 10^{-4}$ | $3.197 \cdot 10^{-3}$ |
|  | $10^{-7}$ | $1.222 \cdot 10^{-5}$ | $2.586$ | $6.829 \cdot 10^{-1}$ | $3.89 \cdot 10^{-3}$ | $2.115 \cdot 10^{-2}$ |
|  | $10^{-8}$ | $3.84 \cdot 10^{-4}$ | $2.618$ | $6.897 \cdot 10^{-2}$ | $1.416 \cdot 10^{-2}$ | $7.695 \cdot 10^{-2}$ |
|  | $10^{-9}$ | $1.213 \cdot 10^{-2}$ | $2.621$ | $6.904 \cdot 10^{-3}$ | $4.547 \cdot 10^{-2}$ | $2.442 \cdot 10^{-1}$ |
|  | $10^{-10}$ | $3.837 \cdot 10^{-1}$ | $2.621$ | $6.905 \cdot 10^{-4}$ | $1.44 \cdot 10^{-1}$ | $5.335 \cdot 10^{-1}$ |

T 5

|  | $R_0$ | $\omega_0$ | $X_0$ | $\beta_{th0}$ | $\frac{R\omega_0}{u}$ | $e_{aZ}^2(R_0)$ | $e_{aT}^2(R_0)$ |
|---|---|---|---|---|---|---|---|
| $p_0 = 10000$ Pa | $10^{-1}$ | $5.571 \cdot 10^{1}$ | $5.278 \cdot 10^{2}$ | $6.295 \cdot 10^{-2}$ | $3.842 \cdot 10^{-3}$ | $1.841 \cdot 10^{-34}$ | $1.001 \cdot 10^{-33}$ |
| $\mu = 0.0014$ Pa s | $10^{-2}$ | $5.573 \cdot 10^{2}$ | $1.669 \cdot 10^{2}$ | $1.965$ | $3.843 \cdot 10^{-3}$ | $1.031 \cdot 10^{-34}$ | $5.607 \cdot 10^{-34}$ |
|  | $10^{-3}$ | $5.591 \cdot 10^{3}$ | $5.287 \cdot 10^{1}$ | $5.969 \cdot 10^{1}$ | $3.856 \cdot 10^{-3}$ | $4.477 \cdot 10^{-35}$ | $2.434 \cdot 10^{-34}$ |
|  | $10^{-4}$ | $5.772 \cdot 10^{4}$ | $1.699 \cdot 10^{1}$ | $1.68 \cdot 10^{3}$ | $3.981 \cdot 10^{-3}$ | $1.918 \cdot 10^{-35}$ | $1.043 \cdot 10^{-34}$ |
|  | $10^{-5}$ | $7.343 \cdot 10^{5}$ | $6.059$ | $4.084 \cdot 10^{4}$ | $5.064 \cdot 10^{-3}$ | $1.601 \cdot 10^{-35}$ | $8.705 \cdot 10^{-35}$ |
|  | $10^{-6}$ | $1.612 \cdot 10^{7}$ | $2.839$ | $5.507 \cdot 10^{5}$ | $1.112 \cdot 10^{-2}$ | $8.954 \cdot 10^{-35}$ | $4.868 \cdot 10^{-34}$ |
|  | $10^{-7}$ | $4.816 \cdot 10^{8}$ | $1.552$ | $5.462 \cdot 10^{6}$ | $3.322 \cdot 10^{-2}$ | $5.923 \cdot 10^{-34}$ | $3.22 \cdot 10^{-33}$ |
|  | $10^{-8}$ | $1.514 \cdot 10^{10}$ | $8.7 \cdot 10^{-1}$ | $5.451 \cdot 10^{7}$ | $1.044 \cdot 10^{-1}$ | $2.156 \cdot 10^{-33}$ | $1.172 \cdot 10^{-32}$ |
|  | $10^{-9}$ | $4.784 \cdot 10^{11}$ | $4.891 \cdot 10^{-1}$ | $5.45 \cdot 10^{8}$ | $3.3 \cdot 10^{-1}$ | $6.924 \cdot 10^{-33}$ | $3.719 \cdot 10^{-32}$ |
|  | $10^{-10}$ | $1.513 \cdot 10^{13}$ | $2.75 \cdot 10^{-1}$ | $5.449 \cdot 10^{9}$ | $1.043$ | $2.193 \cdot 10^{-32}$ | $8.123 \cdot 10^{-32}$ |

T 6

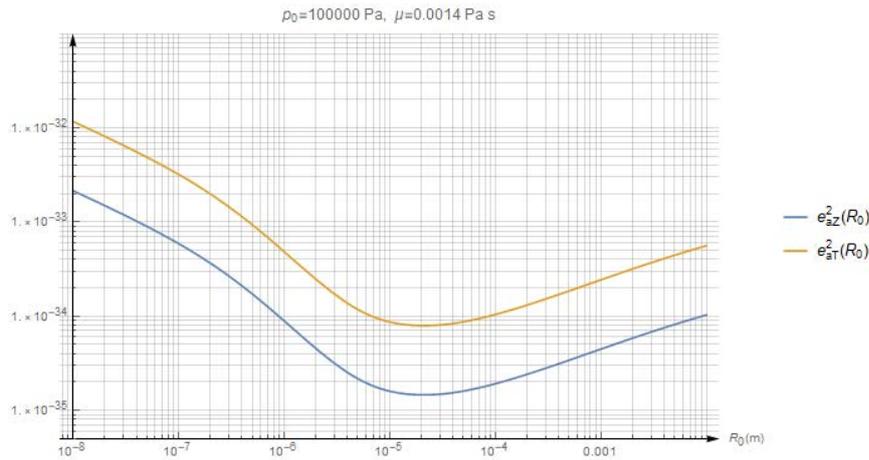

G 3

According to graph G 3, the acoustic charges, the fine acoustic constants and implicitly the acoustic forces have a minimum value for the radius $R_m \cong 2,109 \times 10^{-5}$ m. The square of the acoustic charges corresponding to this radius are: $e_{aZ}^2 \cong 1,466 \cdot 10^{-35}$ Nm$^2$ and $e_{aT}^2 \cong 7,975 \cdot 10^{-35}$ Nm$^2$.

## 2.3. Bubbles in Liquid Helium

Helium behaves as a liquid for temperature values higher than the critical temperature $T = 2.167$ K. Relevant parameters for liquid Helium ($T = 2.5$ K) are: $p_0 = 1.023 \cdot 10^4$ Pa static pressure in liquid, $\mu = 2.4 \cdot 10^{-6}$ Pas dynamic viscosity, $\sigma = 2.62 \cdot 10^{-4}$ Nm$^{-1}$ surface tension, $u = 2.22 \cdot 10^2$ ms$^{-1}$ sound velocity, $\gamma = 1.4$ polytropic exponents of the air in the bubble, $\rho = 1.448 \cdot 10^2$ kgm$^{-3}$ mass density, $\chi = 4.53 \cdot 10^{-8}$ m$^2$s$^{-1}$ self-diffusion coefficient and the universal constants $k = 1.381 \times 10^{-23}$ J·K$^{-1}$, $\hbar = 1.05 \times 10^{-34}$ Js [11, 12]. The results are presented in tables T 7, T 8 and graph G 4.

|  | $R_0$ | $\frac{\hbar\omega_0}{kT}$ | $\frac{4u\mu}{\rho R^3 \omega_0^2}$ | $\frac{2u\beta_{th0}}{R\omega_0^2}$ | $\alpha_{aZ} = \frac{e_{aZ}^2(R_0)}{\hbar u}$ | $\alpha_{aT} = \frac{e_{aT}^2(R_0)}{\hbar u}$ |
|---|---|---|---|---|---|---|
| $p_0 = 10230.\,Pa$ | $10^{-1}$ | $5.72 \cdot 10^{-10}$ | $4.34 \cdot 10^{-7}$ | $2.592 \cdot 10^{-3}$ | $4.222 \cdot 10^{-2}$ | $2.295 \cdot 10^{-1}$ |
| $\mu = 0.0000025\,Pa\,s$ | $10^{-2}$ | $5.72 \cdot 10^{-9}$ | $4.34 \cdot 10^{-6}$ | $8.189 \cdot 10^{-3}$ | $4.199 \cdot 10^{-2}$ | $2.283 \cdot 10^{-1}$ |
|  | $10^{-3}$ | $5.72 \cdot 10^{-8}$ | $4.34 \cdot 10^{-5}$ | $2.582 \cdot 10^{-2}$ | $4.126 \cdot 10^{-2}$ | $2.243 \cdot 10^{-1}$ |
|  | $10^{-4}$ | $5.721 \cdot 10^{-7}$ | $4.338 \cdot 10^{-4}$ | $8.085 \cdot 10^{-2}$ | $3.916 \cdot 10^{-2}$ | $2.129 \cdot 10^{-1}$ |
|  | $10^{-5}$ | $5.732 \cdot 10^{-6}$ | $4.322 \cdot 10^{-3}$ | $2.474 \cdot 10^{-1}$ | $3.389 \cdot 10^{-2}$ | $1.842 \cdot 10^{-1}$ |
|  | $10^{-6}$ | $5.836 \cdot 10^{-5}$ | $4.169 \cdot 10^{-2}$ | $6.934 \cdot 10^{-1}$ | $2.489 \cdot 10^{-2}$ | $1.353 \cdot 10^{-1}$ |
|  | $10^{-7}$ | $6.792 \cdot 10^{-4}$ | $3.079 \cdot 10^{-1}$ | $1.332$ | $1.904 \cdot 10^{-2}$ | $1.034 \cdot 10^{-1}$ |
|  | $10^{-8}$ | $1.292 \cdot 10^{-2}$ | $8.514 \cdot 10^{-1}$ | $8.289 \cdot 10^{-1}$ | $3.566 \cdot 10^{-2}$ | $1.914 \cdot 10^{-1}$ |
|  | $10^{-9}$ | $3.706 \cdot 10^{-1}$ | $1.034$ | $1.244 \cdot 10^{-1}$ | $1.271 \cdot 10^{-1}$ | $4.769 \cdot 10^{-1}$ |
|  | $10^{-10}$ | $1.159 \cdot 10^{1}$ | $1.057$ | $1.3 \cdot 10^{-2}$ | $4.145 \cdot 10^{-1}$ | $2.079 \cdot 10^{-5}$ |

T 7

|  | $R_0$ | $\omega_0$ | $X_0$ | $\beta_{th0}$ | $\frac{R\omega_0}{u}$ | $e_{aZ}^2(R_0)$ | $e_{aT}^2(R_0)$ |
|---|---|---|---|---|---|---|---|
| $p_0 = 10230.\,Pa$ | $10^{-1}$ | $1.879 \cdot 10^{2}$ | $9.109 \cdot 10^{3}$ | $2.062 \cdot 10^{-2}$ | $8.466 \cdot 10^{-2}$ | $9.842 \cdot 10^{-34}$ | $5.351 \cdot 10^{-33}$ |
| $\mu = 0.0000025\,Pa\,s$ | $10^{-2}$ | $1.879 \cdot 10^{3}$ | $2.881 \cdot 10^{3}$ | $6.516 \cdot 10^{-1}$ | $8.466 \cdot 10^{-2}$ | $9.787 \cdot 10^{-34}$ | $5.321 \cdot 10^{-33}$ |
|  | $10^{-3}$ | $1.88 \cdot 10^{4}$ | $9.109 \cdot 10^{2}$ | $2.054 \cdot 10^{1}$ | $8.466 \cdot 10^{-2}$ | $9.619 \cdot 10^{-34}$ | $5.229 \cdot 10^{-33}$ |
|  | $10^{-4}$ | $1.88 \cdot 10^{5}$ | $2.881 \cdot 10^{2}$ | $6.435 \cdot 10^{2}$ | $8.468 \cdot 10^{-2}$ | $9.127 \cdot 10^{-34}$ | $4.962 \cdot 10^{-33}$ |
|  | $10^{-5}$ | $1.883 \cdot 10^{6}$ | $9.119 \cdot 10^{1}$ | $1.977 \cdot 10^{4}$ | $8.483 \cdot 10^{-2}$ | $7.899 \cdot 10^{-34}$ | $4.294 \cdot 10^{-33}$ |
|  | $10^{-6}$ | $1.918 \cdot 10^{7}$ | $2.91 \cdot 10^{1}$ | $5.743 \cdot 10^{5}$ | $8.638 \cdot 10^{-2}$ | $5.802 \cdot 10^{-34}$ | $3.154 \cdot 10^{-33}$ |
|  | $10^{-7}$ | $2.232 \cdot 10^{8}$ | $9.926$ | $1.494 \cdot 10^{7}$ | $1.005 \cdot 10^{-1}$ | $4.438 \cdot 10^{-34}$ | $2.411 \cdot 10^{-33}$ |
|  | $10^{-8}$ | $4.244 \cdot 10^{9}$ | $4.328$ | $3.362 \cdot 10^{8}$ | $1.912 \cdot 10^{-1}$ | $8.312 \cdot 10^{-34}$ | $4.461 \cdot 10^{-33}$ |
|  | $10^{-9}$ | $1.218 \cdot 10^{11}$ | $2.319$ | $4.156 \cdot 10^{9}$ | $5.485 \cdot 10^{-1}$ | $2.962 \cdot 10^{-33}$ | $1.112 \cdot 10^{-32}$ |
|  | $10^{-10}$ | $3.809 \cdot 10^{12}$ | $1.297$ | $4.25 \cdot 10^{10}$ | $1.716$ | $9.663 \cdot 10^{-33}$ | $4.847 \cdot 10^{-37}$ |

T 8

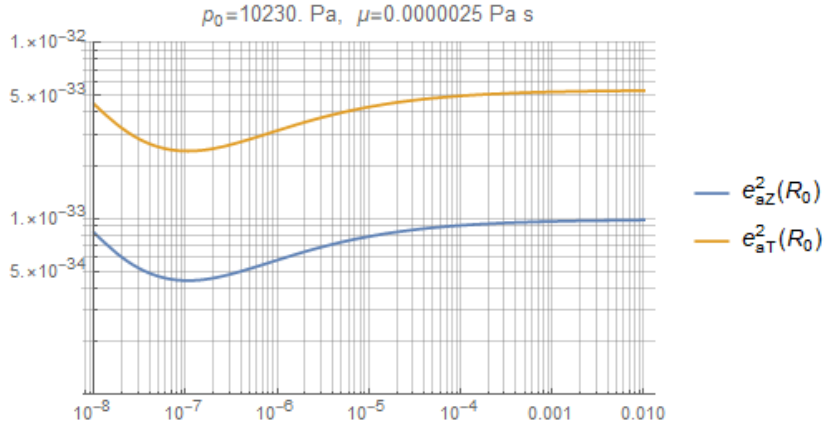

G 4.

Liquid helium, at $T = 2.5\,K$, influences the size of the forces and the square of the acoustic charge, similar to the other investigated liquids. According to the G4 graph, the acoustic charges, the fine acoustic constants and implicitly the acoustic forces have a minimum value for the radii: $R_{HeZ} \cong 1,0614 \times 10^{-7}\,m$ and $R_{HeT} \cong 1,0573 \times 10^{-7}\,m$. The square of the acoustic charges corresponding to these radii are: $e_{aZ}^2 \cong 4,436 \cdot 10^{-34}\,Nm^2$ and $e_{aT}^2 \cong 2,4105 \cdot 10^{-35}\,Nm^2$. Unlike water and mercury, the minimum for the square of the charge, for liquid helium, is obtained at different values of the radius for the two acoustic radiation backgrounds.

## 2.4. Bubbles in Superfluid Helium

Helium behaves as a superfluid for temperature values in the range $T \in [1.5 - 2.16]\,K$. Superfluid helium parameters at $T = 1.5\,K$, relevant for the phenomenon of interaction between bubbles, are: $p_0 = 471\,Pa$ static pressure in liquid, $\mu = 2.6 \cdot 10^{-6}\,Pa\,s$ dynamic viscosity, $\sigma = 3.2 \cdot 10^{-4}\,Nm^{-1}$ surface tension, $u = 234\,ms^{-1}$ sound velocity, $\gamma = 5/3$ polytropic exponents of the vapors in the bubble, $\rho = 145\,kgm^{-3}$ mass density, $\chi = 7 \cdot 10^{6}\,m^2 s^{-1}$ self-diffusion coefficient for vapors and the universal constants $k = 1.381 \times 10^{-23}\,JK^{-1}$, $\hbar = 1.05 \times 10^{-34}\,Js$ [11, 12]. We mention that the superfluid coefficient was calculated from the experimental results for superfluid thermal conductivity $\kappa$ and specific heat capacity $C_p$ ($\chi = \kappa/(\rho C_p)$), which we found, each, from only one academic source [13, 14]. The results are presented in tables T 9, T 10 and graph G 5.

| | $R_0$ | $\dfrac{\hbar\omega_0}{kT}$ | $\dfrac{4u\mu}{\rho R^3 \omega_0^2}$ | $\dfrac{2u\beta_{th0}}{R\omega_0^2}$ | $\alpha_{aZ} = \dfrac{e_{aZ}^2(R_0)}{\hbar u}$ | $\alpha_{aT} = \dfrac{e_{aT}^2(R_0)}{\hbar u}$ |
|---|---|---|---|---|---|---|
| $p_0 = 471.\,\text{Pa}$ | $10^{-1}$ | $2.044 \cdot 10^{-10}$ | $1.033 \cdot 10^{-5}$ | $-9.288 \cdot 10^{-2}$ | $9.493 \cdot 10^{-3}$ | $5.161 \cdot 10^{-2}$ |
| $\mu = 0.0000026\,\text{Pa s}$ | $10^{-2}$ | $2.044 \cdot 10^{-9}$ | $1.033 \cdot 10^{-4}$ | $1.154 \cdot 10^{2}$ | $7.395 \cdot 10^{-5}$ | $4.021 \cdot 10^{-4}$ |
| | $10^{-3}$ | $2.045 \cdot 10^{-8}$ | $1.032 \cdot 10^{-3}$ | $-1.011 \cdot 10^{4}$ | $-8.524 \cdot 10^{-7}$ | $-4.634 \cdot 10^{-6}$ |
| | $10^{-4}$ | $2.055 \cdot 10^{-7}$ | $1.022 \cdot 10^{-2}$ | $-2.54 \cdot 10^{5}$ | $-3.409 \cdot 10^{-8}$ | $-1.853 \cdot 10^{-7}$ |
| | $10^{-5}$ | $2.152 \cdot 10^{-6}$ | $9.321 \cdot 10^{-2}$ | $6.571 \cdot 10^{7}$ | $1.38 \cdot 10^{-10}$ | $7.502 \cdot 10^{-10}$ |
| | $10^{-6}$ | $2.953 \cdot 10^{-5}$ | $4.951 \cdot 10^{-1}$ | $-3.775 \cdot 10^{9}$ | $-3.295 \cdot 10^{-12}$ | $-1.791 \cdot 10^{-11}$ |
| | $10^{-7}$ | $7.043 \cdot 10^{-4}$ | $8.705 \cdot 10^{-1}$ | $8.092 \cdot 10^{9}$ | $3.667 \cdot 10^{-12}$ | $1.992 \cdot 10^{-11}$ |
| | $10^{-8}$ | $2.141 \cdot 10^{-2}$ | $9.42 \cdot 10^{-1}$ | $-7.195 \cdot 10^{10}$ | $-1.254 \cdot 10^{-12}$ | $-6.671 \cdot 10^{-12}$ |
| | $10^{-9}$ | $6.743 \cdot 10^{-1}$ | $9.498 \cdot 10^{-1}$ | $-2.901 \cdot 10^{11}$ | $-9.792 \cdot 10^{-13}$ | $-2.712 \cdot 10^{-12}$ |
| | $10^{-10}$ | $2.131 \cdot 10^{1}$ | $9.505 \cdot 10^{-1}$ | $1.405 \cdot 10^{11}$ | $6.389 \cdot 10^{-12}$ | $1.923 \cdot 10^{-20}$ |

T9

| | $R_0$ | $\omega_0$ | $X_0$ | $\beta_{th0}$ | $\dfrac{R\omega_0}{u}$ | $e_{aZ}^2(R_0)$ | $e_{aT}^2(R_0)$ |
|---|---|---|---|---|---|---|---|
| $p_0 = 471.\,\text{Pa}$ | $10^{-1}$ | $4.03 \cdot 10^{1}$ | $3.393 \cdot 10^{-4}$ | $-3.223 \cdot 10^{-2}$ | $1.722 \cdot 10^{-2}$ | $2.332 \cdot 10^{-34}$ | $1.268 \cdot 10^{-33}$ |
| $\mu = 0.0000026\,\text{Pa s}$ | $10^{-2}$ | $4.03 \cdot 10^{2}$ | $1.073 \cdot 10^{-4}$ | $4.007 \cdot 10^{2}$ | $1.722 \cdot 10^{-2}$ | $1.817 \cdot 10^{-36}$ | $9.879 \cdot 10^{-36}$ |
| | $10^{-3}$ | $4.032 \cdot 10^{3}$ | $3.394 \cdot 10^{-5}$ | $-3.512 \cdot 10^{5}$ | $1.723 \cdot 10^{-2}$ | $-2.094 \cdot 10^{-38}$ | $-1.139 \cdot 10^{-37}$ |
| | $10^{-4}$ | $4.052 \cdot 10^{4}$ | $1.076 \cdot 10^{-5}$ | $-8.911 \cdot 10^{7}$ | $1.732 \cdot 10^{-2}$ | $-8.375 \cdot 10^{-40}$ | $-4.553 \cdot 10^{-39}$ |
| | $10^{-5}$ | $4.243 \cdot 10^{5}$ | $3.482 \cdot 10^{-6}$ | $2.528 \cdot 10^{11}$ | $1.813 \cdot 10^{-2}$ | $3.39 \cdot 10^{-42}$ | $1.843 \cdot 10^{-41}$ |
| | $10^{-6}$ | $5.822 \cdot 10^{6}$ | $1.29 \cdot 10^{-6}$ | $-2.734 \cdot 10^{14}$ | $2.488 \cdot 10^{-2}$ | $-8.096 \cdot 10^{-44}$ | $-4.401 \cdot 10^{-43}$ |
| | $10^{-7}$ | $1.388 \cdot 10^{8}$ | $6.299 \cdot 10^{-7}$ | $3.333 \cdot 10^{16}$ | $5.934 \cdot 10^{-2}$ | $9.009 \cdot 10^{-44}$ | $4.894 \cdot 10^{-43}$ |
| | $10^{-8}$ | $4.221 \cdot 10^{9}$ | $3.473 \cdot 10^{-7}$ | $-2.739 \cdot 10^{19}$ | $1.804 \cdot 10^{-1}$ | $-3.08 \cdot 10^{-44}$ | $-1.639 \cdot 10^{-43}$ |
| | $10^{-9}$ | $1.329 \cdot 10^{11}$ | $1.949 \cdot 10^{-7}$ | $-1.095 \cdot 10^{22}$ | $5.681 \cdot 10^{-1}$ | $-2.406 \cdot 10^{-44}$ | $-6.664 \cdot 10^{-44}$ |
| | $10^{-10}$ | $4.202 \cdot 10^{12}$ | $1.096 \cdot 10^{-7}$ | $5.302 \cdot 10^{23}$ | $1.796$ | $1.57 \cdot 10^{-43}$ | $4.725 \cdot 10^{-52}$ |

T10

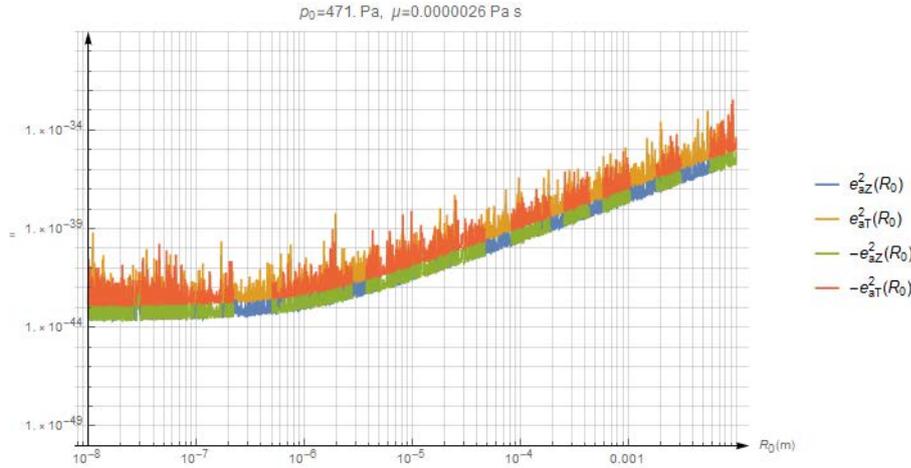

G 5

The tables T 9, T 10 and graphs G 5 show the variation of the square of the acoustic charge from positive to negative values for small variations in the bubble radius. The change in sign indicates a change in sense of the acoustic forces from attractive to repulsive forces. This phenomenon occurs as a result of the fact that the thermal coefficient $\beta_{th0}$ is dependent on the trigonometric functions $\sin\left[R_0\left(2\omega_0/\chi\right)^{1/2}\right]$ and $\cos\left[R_0\left(2\omega_0/\chi\right)^{1/2}\right]$, which in turn depend on the radius and the coefficient $\chi$, which for the superfluid is very large. For certain values of the radius, the ratio $2u\beta_{th0}/(R_0\omega_0^2)$ is negative and much larger, in absolute value, than $1 + 4\mu u/(\rho R_0^3 \omega_0^2)$. It follows that $1 + 4\mu u/(\rho R_0^3 \omega_0^2) + 2u\beta_{th0}/(R_0\omega_0^2)$ it is negative and the forces, according to Eq. (1), changes its sense. According to graph G 5, there are ranges of radii (marked with blue and yellow colors) for which the forces (the square of the acoustic charge is positive) are attractive and ranges (marked with the colors green and orange) for which the forces (the square of the acoustic charge is negative) repulsive.

We find the same dependence for helium at temperature $T = 2K$. Superfluid helium parameters at $T = 2K$, relevant for the phenomenon of interaction between bubbles, are: $p_0 = 3130 \, Pa$ static pressure in liquid, $\mu = 10^{-6} \, Pa \, s$ dynamic viscosity, $\sigma = 3 \cdot 10^{-4} \, Nm^{-1}$ surface tension, $u = 226 \, ms^{-1}$ sound velocity, $\gamma = 5/3$ polytropic exponents of the vapors in the bubble, $\rho = 145 \, kgm^{-3}$ mass density, $\chi = 2 \cdot 10^7 \, m^2 s^{-1}$ self-diffusion coefficient for vapors [11-14] and the universal constants $k = 1.381 \times 10^{-23} \, JK^{-1}$, $\hbar = 1.05 \times 10^{-34} \, Js$. The results are presented in tables T 11, T 12 and graph G 6.

|  | $R_0$ | $\dfrac{\hbar\omega_0}{kT}$ | $\dfrac{4u\mu}{\rho R^3 \omega_0^2}$ | $\dfrac{2u\beta_{th0}}{R\omega_0^2}$ | $\alpha_{aZ} = \dfrac{e_{aZ}^2(R_0)}{\hbar u}$ | $\alpha_{aT} = \dfrac{e_{aT}^2(R_0)}{\hbar u}$ |
|---|---|---|---|---|---|---|
| $p_0 = 3130. \, Pa$ | $10^{-1}$ | $3.952 \cdot 10^{-10}$ | $8.491 \cdot 10^{-7}$ | $-5.136 \cdot 10^{-1}$ | $4.726 \cdot 10^{-2}$ | $2.569 \cdot 10^{-1}$ |
| $\mu = 0.00000147 \, Pa \, s$ | $10^{-2}$ | $3.952 \cdot 10^{-9}$ | $8.491 \cdot 10^{-6}$ | $-3.419 \cdot 10^{1}$ | $-6.926 \cdot 10^{-4}$ | $-3.765 \cdot 10^{-3}$ |
|  | $10^{-3}$ | $3.953 \cdot 10^{-8}$ | $8.49 \cdot 10^{-5}$ | $1.992 \cdot 10^{3}$ | $1.154 \cdot 10^{-5}$ | $6.272 \cdot 10^{-5}$ |
|  | $10^{-4}$ | $3.955 \cdot 10^{-7}$ | $8.478 \cdot 10^{-4}$ | $4.966 \cdot 10^{5}$ | $4.632 \cdot 10^{-8}$ | $2.518 \cdot 10^{-7}$ |
|  | $10^{-5}$ | $3.983 \cdot 10^{-6}$ | $8.363 \cdot 10^{-3}$ | $-2.876 \cdot 10^{7}$ | $-8.051 \cdot 10^{-10}$ | $-4.377 \cdot 10^{-9}$ |
|  | $10^{-6}$ | $4.245 \cdot 10^{-5}$ | $7.362 \cdot 10^{-2}$ | $2.159 \cdot 10^{9}$ | $1.143 \cdot 10^{-11}$ | $6.215 \cdot 10^{-11}$ |
|  | $10^{-7}$ | $6.291 \cdot 10^{-4}$ | $3.352 \cdot 10^{-1}$ | $6.056 \cdot 10^{10}$ | $6.041 \cdot 10^{-13}$ | $3.282 \cdot 10^{-12}$ |
|  | $10^{-8}$ | $1.597 \cdot 10^{-2}$ | $5.198 \cdot 10^{-1}$ | $-6.457 \cdot 10^{11}$ | $-1.439 \cdot 10^{-13}$ | $-7.698 \cdot 10^{-13}$ |
|  | $10^{-9}$ | $4.91 \cdot 10^{-1}$ | $5.501 \cdot 10^{-1}$ | $7.308 \cdot 10^{11}$ | $3.907 \cdot 10^{-13}$ | $1.3 \cdot 10^{-12}$ |
|  | $10^{-10}$ | $1.548 \cdot 10^{1}$ | $5.533 \cdot 10^{-1}$ | $6.135 \cdot 10^{12}$ | $1.468 \cdot 10^{-13}$ | $1.506 \cdot 10^{-19}$ |

T 11

|  | $R_0$ | $\omega_0$ | $X_0$ | $\beta_{th0}$ | $\dfrac{R\omega_0}{u}$ | $e_{aZ}^2(R_0)$ | $e_{aT}^2(R_0)$ |
|---|---|---|---|---|---|---|---|
| $p_0 = 3130. \, Pa$ | $10^{-1}$ | $1.039 \cdot 10^{2}$ | $3.223 \cdot 10^{-4}$ | $-1.226$ | $4.597 \cdot 10^{-2}$ | $1.121 \cdot 10^{-33}$ | $6.097 \cdot 10^{-33}$ |
| $\mu = 0.00000147 \, Pa \, s$ | $10^{-2}$ | $1.039 \cdot 10^{3}$ | $1.019 \cdot 10^{-4}$ | $-8.163 \cdot 10^{2}$ | $4.597 \cdot 10^{-2}$ | $-1.644 \cdot 10^{-35}$ | $-8.935 \cdot 10^{-35}$ |
|  | $10^{-3}$ | $1.039 \cdot 10^{4}$ | $3.223 \cdot 10^{-5}$ | $4.756 \cdot 10^{5}$ | $4.597 \cdot 10^{-2}$ | $2.738 \cdot 10^{-37}$ | $1.488 \cdot 10^{-36}$ |
|  | $10^{-4}$ | $1.04 \cdot 10^{5}$ | $1.02 \cdot 10^{-5}$ | $1.188 \cdot 10^{9}$ | $4.6 \cdot 10^{-2}$ | $1.099 \cdot 10^{-39}$ | $5.976 \cdot 10^{-39}$ |
|  | $10^{-5}$ | $1.047 \cdot 10^{6}$ | $3.235 \cdot 10^{-6}$ | $-6.974 \cdot 10^{11}$ | $4.632 \cdot 10^{-2}$ | $-1.911 \cdot 10^{-41}$ | $-1.039 \cdot 10^{-40}$ |
|  | $10^{-6}$ | $1.116 \cdot 10^{7}$ | $1.056 \cdot 10^{-6}$ | $5.947 \cdot 10^{14}$ | $4.937 \cdot 10^{-2}$ | $2.713 \cdot 10^{-43}$ | $1.475 \cdot 10^{-42}$ |
|  | $10^{-7}$ | $1.654 \cdot 10^{8}$ | $4.066 \cdot 10^{-7}$ | $3.664 \cdot 10^{17}$ | $7.317 \cdot 10^{-2}$ | $1.434 \cdot 10^{-44}$ | $7.789 \cdot 10^{-44}$ |
|  | $10^{-8}$ | $4.199 \cdot 10^{9}$ | $2.049 \cdot 10^{-7}$ | $-2.519 \cdot 10^{20}$ | $1.858 \cdot 10^{-1}$ | $-3.414 \cdot 10^{-45}$ | $-1.827 \cdot 10^{-44}$ |
|  | $10^{-9}$ | $1.291 \cdot 10^{11}$ | $1.136 \cdot 10^{-7}$ | $2.694 \cdot 10^{22}$ | $5.711 \cdot 10^{-1}$ | $9.272 \cdot 10^{-45}$ | $3.085 \cdot 10^{-44}$ |
|  | $10^{-10}$ | $4.07 \cdot 10^{12}$ | $6.379 \cdot 10^{-8}$ | $2.248 \cdot 10^{25}$ | $1.801$ | $3.483 \cdot 10^{-45}$ | $3.575 \cdot 10^{-51}$ |

T12

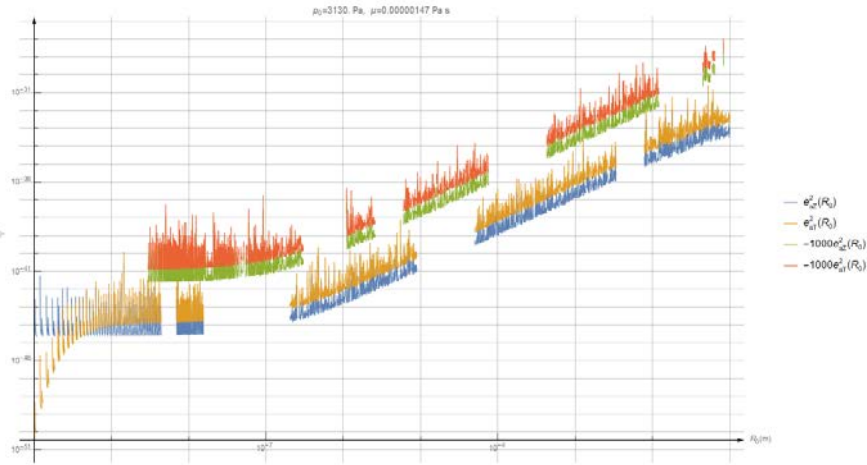

G 6

According to graph G 6, there are ranges of radii (marked with blue and yellow colors) for which the forces (the square of the acoustic charge is positive) are attractive and ranges (marked with the colors green and orange) for which the forces (the square of the acoustic charge is negative) repulsive. The separation of the curves for which the square of the acoustic charge is negative, by amplifying with the factor $10^3$, allows the observation of another characteristic of the dependence on the bubbles radii, namely the fact that the ranges of radii for which the square of the acoustic charge is positive partially overlap the domains in which the square of the acoustic charge is negative. It follows that in the overlapping areas, the square of the acoustic charge oscillates between positive and negative values.

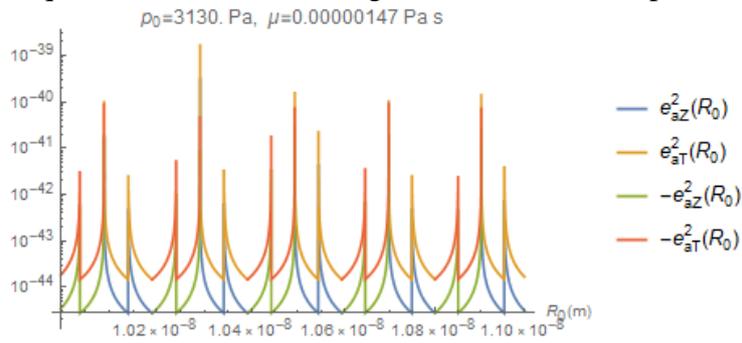

G 7

Graph G 7 shows the dependence of the square of the acoustic charge on the size of the bubble radius in the range $[10^{-8} - 1.1 \cdot 10^{-8}]$ m, to highlight the oscillations between positive and negative values.

In order to better understand the dependence of the square of the acoustic charge on the radii of the bubbles, we also represented the thermal coefficient $\beta_{th0}$ as a function of the radius, according to graphs G 8 - G 11. Graph 8 shows the dependence of the coefficient $\beta_{th0}$ for radii of the order $10^{-10}$ m.

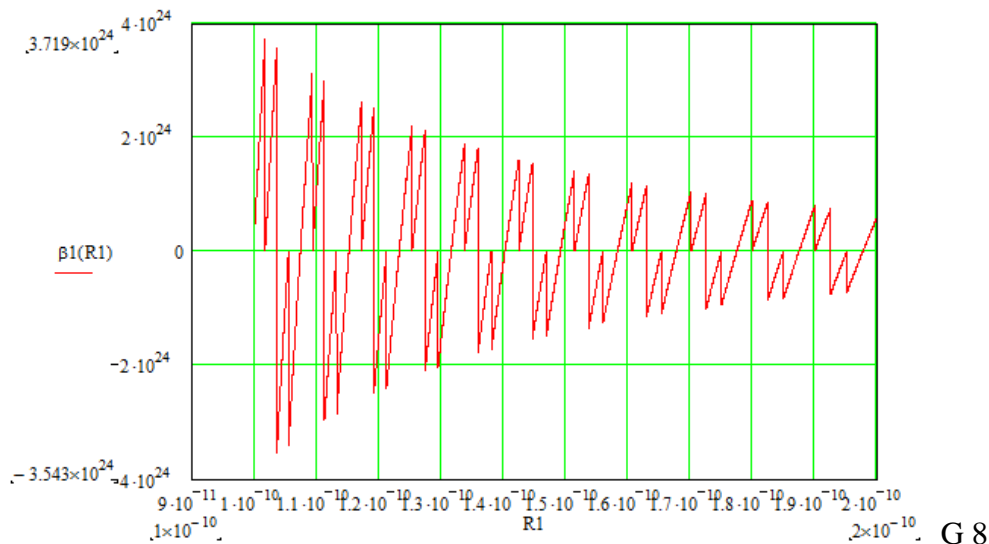

G 8

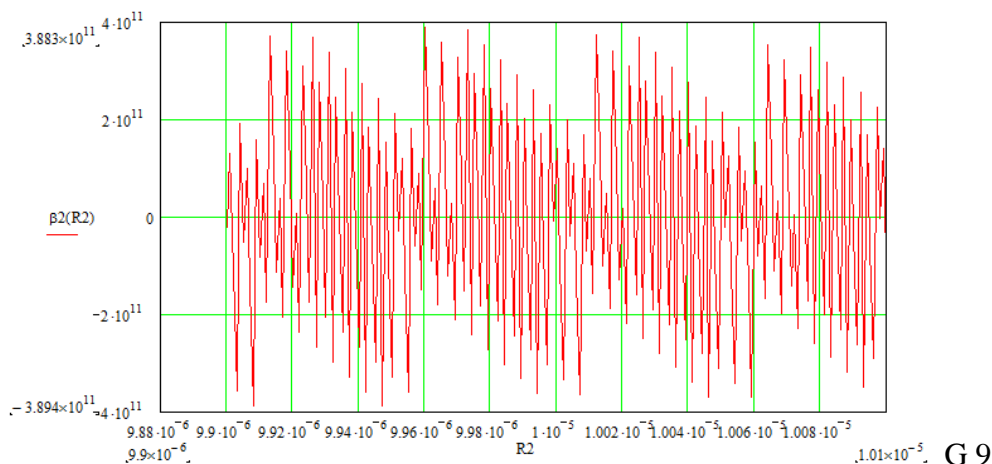

G 9

Graph 9 shows the dependence of the coefficient $\beta_{th0}$ for radii of order $10^{-5}$ m.

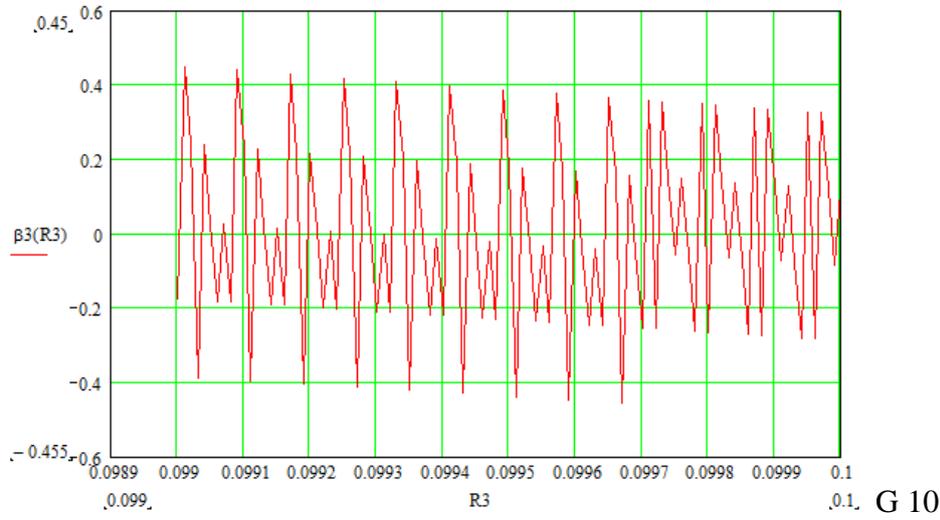

Graph 10 shows the dependence of the coefficient $\beta_{th0}$ for radii of order $10^{-1}$ m.

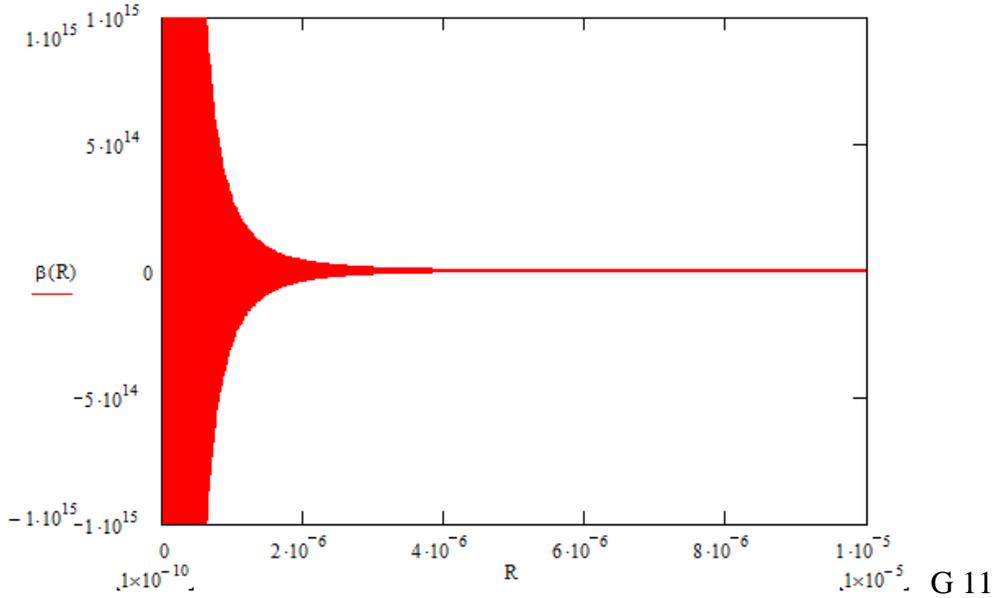

Graph 11 shows the dependence of the coefficient $\beta_{th0}$ for radii in the interval $R_0 \in [0 - 10^{-5}]$ m. For small radii, i.e. at the physical limit of bubble radii in the fluid, $R_0 \in [10^{-10} - 10^{-7}]$ m, the oscillation phenomenon is amplified by the ratio $\omega_0/X_0 = (1/R_0)(\omega_0 \chi/2)^{1/2}$ which is very large.

## 3. Comparison of the scattering-scattering force with the scattering-absorption force

According to the paper [3], the scattering-scattering forces and the scattering-absorption forces, in the radiation background, have the expressions:

$$F_{BssT} = \left(\frac{-\hbar u}{r^2}\right) \frac{2\beta_{s0}}{\omega_0 \left(\exp\frac{\hbar \omega_0}{kT} - 1\right)\left(1 + \beta_{\mu 0}/\beta_{s0} + \beta_{th 0}/\beta_{s0}\right)}, \qquad (10)$$

$$F_{BssZ} = \left(\frac{-\hbar u}{r^2}\right) \frac{\beta_{s0}}{\omega_0 \left(1 + \beta_{\mu 0}/\beta_{s0} + \beta_{th 0}/\beta_{s0}\right)}, \tag{11}$$

$$F_{BsaT} = \left(\frac{-\hbar u}{r^2}\right) \frac{\beta_{\mu 0} + \beta_{th 0}}{\omega_0 \left(\exp\frac{\hbar \omega_0}{kT} - 1\right)\left(1 + \beta_{\mu 0}/\beta_{s0} + \beta_{th 0}/\beta_{s0}\right)^3}, \tag{12}$$

$$F_{BsaZ} = \left(\frac{-\hbar u}{r^2}\right) \frac{\beta_{\mu 0} + \beta_{th 0}}{2\omega_0 \left(1 + \beta_{\mu 0}/\beta_{s0} + \beta_{th 0}/\beta_{s0}\right)^3}. \tag{13}$$

Comparing them, it follows:

$$f_a = \frac{F_{BsaT}}{F_{BssT}} = \frac{\beta_{\mu 0}/\beta_{s0} + \beta_{th 0}/\beta_{s0}}{\left(1 + \beta_{\mu 0}/\beta_{s0} + \beta_{th 0}/\beta_{s0}\right)^2} = \frac{F_{BsaZ}}{F_{BssZ}}. \tag{14}$$

According to the result of the numerical analysis made in section 2, the ratio of the two types of forces, $f_a$, for bubble radii included in the interval $R_0 = [10^{-1} - 10^{-8}]$m, is less than one. As an example, for $R_0 = 10^{-1}$m, the ratio is: $f_a = 0.42$ for water at $p_0 = 10^5$Pa, $f_a = 0.0006$ for water at $p_0 = 0$Pa, $f_a = 0.23$ for mercury at $p_0 = 10^5$Pa, $f_a = 0.0026$ for liquid helium at $p_0 = 1.023 \cdot 10^5$Pa. For $R_0 = 10^{-5}$m, the ratio is: $f_a = 0.17$ for water at $p_0 = 10^5$Pa, $f_a = 0.16$ for water at $p_0 = 0$Pa, $f_a = 0.04$ for mercury at $p_0 = 10^5$Pa, $f_a = 0.16$ for liquid helium at $p_0 = 1.023 \cdot 10^5$Pa. For $R_0 = 10^{-8}$m, the ratio is: $f_a = 0.07$ for water at $p_0 = 10^5$Pa, $f_a = 0.24$ for water at $p_0 = 0$Pa, $f_a = 0.2$ for mercury at $p_0 = 10^5$Pa, $f_a = 0.23$ for liquid helium at $p_0 = 1.023 \cdot 10^5$Pa.

For superfluid helium, the ratio is much less than one and oscillates between positive and negative values (changing the direction of the forces). For instance: for $R_0 = 10^{-1}$m, the ratio is $f_a = -0.01$ for superfluid helium at temperature $T = 1.5$K and $f_a = -2.2$ for superfluid helium at temperature $T = 2$K; for $R_0 = 10^{-5}$m, the ratio is $f_a = 1.5 \cdot 10^{-8}$ for superfluid helium at temperature $T = 1.5$K and $f_a = -3.5 \cdot 10^{-8}$ for superfluid helium at temperature $T = 2$K; for $R_0 = 10^{-8}$m, the ratio is $f_a = -1.3 \cdot 10^{-11}$ for superfluid helium at temperature $T = 1.5$K and $f_a = -1.6 \cdot 10^{-12}$ for superfluid helium at temperature $T = 2$K. It follows that only for superfluid helium the ratio is very small, in absolute value, but not as small as the ratio of gravitational forces to electrostatic forces, $f_e = F_g/F_e = Gm_e^2/e^2 \cong 10^{-42}$, for the same charged particle (in this case, the electron) in the electromagnetic world. In a future paper we will show that their ratio is very small in the case of forces between two bubbles in a cluster with a large number of bubbles, $N \gg 1$. In this case, the ratio depends on the acoustic Mach number, $N_c \cong N(R_0/R_c)$, which depends on the number $N$ of bubbles in the cluster, the cluster radius, and the bubble radius [8], analogous to the dependence of the ratio $f_e$ on the number of particles in the Universe, the universe radius, and the electron radius, according to of the Edington – Dirac hypothesis ($1/f_e = F_e/F_g \cong 10^{42} \cong R_u/R_e$, $R_u \cong NGm_e/c^2$, $R_e = e^2/(m_e c^2)$) [8, 15-17].

## 4. Conclusions

In this paper, we demonstrated by numerical calculation, that the contribution of the attenuation coefficients $\beta_{s0}$ (acoustic damping coefficient) and $\beta_{a0} = \beta_{\mu 0} + \beta_{th 0}$ (absorbtion damping coefficient) to the size of the interaction forces, implicitly the square of the acoustic charges, between two free bubbles oscillating under the action of a stochastic acoustic background (the thermal background and the CZPF background) depends on the radii of the bubbles and the properties of the liquids in which the bubbles form. For liquids with different properties: water, mercury, liquid helium, it turned out that the attenuation coefficients have a comparable contribution in size, for radii in the range $R_0 = [10^{-1} - 10^{-10}]$ m. This result does not justify the approximation, $\beta_{s0} \gg \beta_{a0} = \beta_{\mu 0} + \beta_{th0}$, analogous to that for the attenuation coefficients for the electromagnetic oscillator [6, 7], made in the papers [1-4].

For superfluid helium ($T \in [1.5 - 2.16]$ K), the results of the calculation of the acoustic force as a function of the bubble radius highlight a strong dependence on the diffusion coefficient $\chi$ that leads to variations between positive and negative values (change of sign and therefore the sense of the forces) of the acoustic forces for values near the radii of the bubbles. This result is explained by the dependence of the thermal absorption coefficient $\beta_{th0}$ on the trigonometric functions sin and cos. Also, the values of the coefficient $\beta_{th0}$ are greater, in absolute value, than the coefficients, $\beta_{s0}$ and $\beta_{\mu 0}$ so their sum can be negative.

In this paper, we also calculated the ratio of the scattering–absorption forces to the scattering–scattering forces in the radiation background $f_a = F_{BsaT}/F_{BssT} = F_{BsaZ}/F_{BssZ}$ in these liquids. Calculations show that these forces are close in magnitude for the same bubble radii. The exception is this ratio calculated for bubbles in superfluid helium, which is much smaller than one.

In conclusion, we can consider that the scattering-scattering forces, identified as electrostatic acoustic forces, and the scattering-absorption forces, identified as gravitational acoustic forces, between two free bubbles, do not differ from each other in magnitude, like the same type of forces in the electromagnetic world. This conclusion is also supported by the expressions of the acoustic section in a radiation background, Eqs. (79) from the paper [1-arxivV2], which do not depend on the damping coefficients $\beta_0 = \beta_{s0} + \beta_{\mu 0} + \beta_{th0}$.